# Anisotropic Klemens model for the thermal conductivity tensor

Bo Jiang, Tao Li, and Zhen Chen*

School of Mechanical Engineering, Southeast University, Nanjing 210096, China

* To whom correspondence should be addressed to: zhenchen@seu.edu.cn

Abstract: With the constraint from Onsager reciprocity relations, here we generalize the Klemens model for phonon-phonon Umklapp scattering from isotropic to anisotropic. Combined with the anisotropic Debye dispersion, this anisotropic Klemens model leads to analytical expressions for heat transfer along both *ab*-plane ($\kappa_{ab}$) and *c*-axis ($\kappa_c$), suitable for both layered and chainlike materials with any anisotropy ratio of the dispersion and the scattering. The model is justified by comparison with experimental $\kappa_{ab}$ and $\kappa_c$ of bulk graphite at high temperatures, as well as the thickness-dependent $\kappa_c$ of graphite thin films at room temperature.

Modeling the thermal conductivity of highly anisotropic materials has been attracting both fundamental and applied interest. Although more advanced and accurate models are existing,[1–4] a simple but insightful phenomenological model, e.g. an anisotropic extension to the classic Callaway[5] or Holland[6] model, is still valuable. Solving the Boltzmann transport equation (BTE) under the relaxation time approximation (RTA), one obtains the thermal conductivity tensor,[7]

$$\kappa_{\alpha\beta} = \sum_{pol} \left[ \sum_{\mathbf{k}} C_{\mathbf{k}} \cdot (\mathbf{v}_{g,\mathbf{k}} \cdot \hat{\alpha}) \cdot (\mathbf{v}_{g,\mathbf{k}} \cdot \hat{\beta}) \cdot \tau_{\mathbf{k}} \right], \quad (1)$$

where the summations run over all wavevectors, $\mathbf{k}$, and polarizations. Here $C_{\mathbf{k}}$, $\mathbf{v}_{g,\mathbf{k}}$, and $\tau_{\mathbf{k}}$ are the mode-wise specific heat, group velocity, and relaxation time, respectively, in which we drop their polarization dependence for clarity. The unit vectors, $\hat{\alpha}$ and $\hat{\beta}$, denote the direction of the heat flux and the temperature gradient, respectively. It is self-evident from Eq. 1 that $\kappa_{\alpha\beta} = \kappa_{\beta\alpha}$, as is required by the Onsager reciprocity relations.[8,9]

Equation 1 highlights two key ingredients in modeling $\kappa_{\alpha\beta}$: the dispersion gives $\mathbf{v}_{g,\mathbf{k}}$ and $C_{\mathbf{k}}$, and the scattering gives $\tau_{\mathbf{k}}$. We previously proposed an anisotropic Debye dispersion to model the thermal boundary conductance,[10] and subsequently the minimum thermal conductivity[11] of anisotropic materials with tetragonal, trigonal, or hexagonal symmetries. These models was applied to various anisotropic materials such as $WSe_2$,[11] $MoS_2$,[12] and polymer fibers,[13] and was recently extended to materials with orthorhombic symmetry, such as black phosphorous.[14]

However, a phenomenological model of anisotropic phonon scattering is still lacking. Even for graphite, the most familiar anisotropic material with well-documented experimental data, there is still inconsistency on whether an isotropic scattering time is adequate to model the directional thermal conductivity. With an isotropic scattering time, one approach considered first a lumped one phonon branch model[15] and then a branch-wise model.[16] An alternative approach proposed that the relaxation time varies with the direction of the temperature gradient ($\nabla T$),[17] which was also taken up in other literature for other anisotropic materials.[14,18] However, with details in the supplementary material,[19] we find that this seemingly-reasonable phenomenology of a $\nabla T$ dependent phonon-phonon scattering also turns out to imply a violation of the Onsager reciprocity relations, i.e. the well-established symmetry of the thermal conductivity tensor, $\kappa_{\alpha\beta} = \kappa_{\beta\alpha}$.

To circumvent this shortcoming, we propose a modified anisotropic scattering model, whereby the locus of next collisions is generalized from spherical to ellipsoidal. Combining the anisotropic Debye dispersion and the anisotropic scattering model, we derive analytical expressions of the directional thermal conductivity at high temperatures, and adapt the equation of phonon radiative transfer to model heat conduction along the cross-plane direction of anisotropic films. This model is justified by comparison with experiments of graphite from literature.

Here we consider the imaginary locus of next possible collisions of a phonon with a specific frequency and polarization. In isotropic materials, the locus is a sphere with a radius of one mean free path (MFP), $\Lambda$. In anisotropic materials, however, the symmetry is broken, and the locus of next collisions could be direction-dependent, i.e. non-spherical. As the most important first generalization of anisotropic scattering, we consider an ellipsoidal surface (Fig. 1a),

$$\frac{r^2}{\Lambda_{ab}^2} + \frac{z^2}{\Lambda_c^2} = 1, \qquad (2)$$

where $\Lambda_{ab}$ and $\Lambda_c$ are the equatorial and polar radius, respectively. This description may also be viewed as a generalized concept of optical[20] or acoustic[21] thickness.

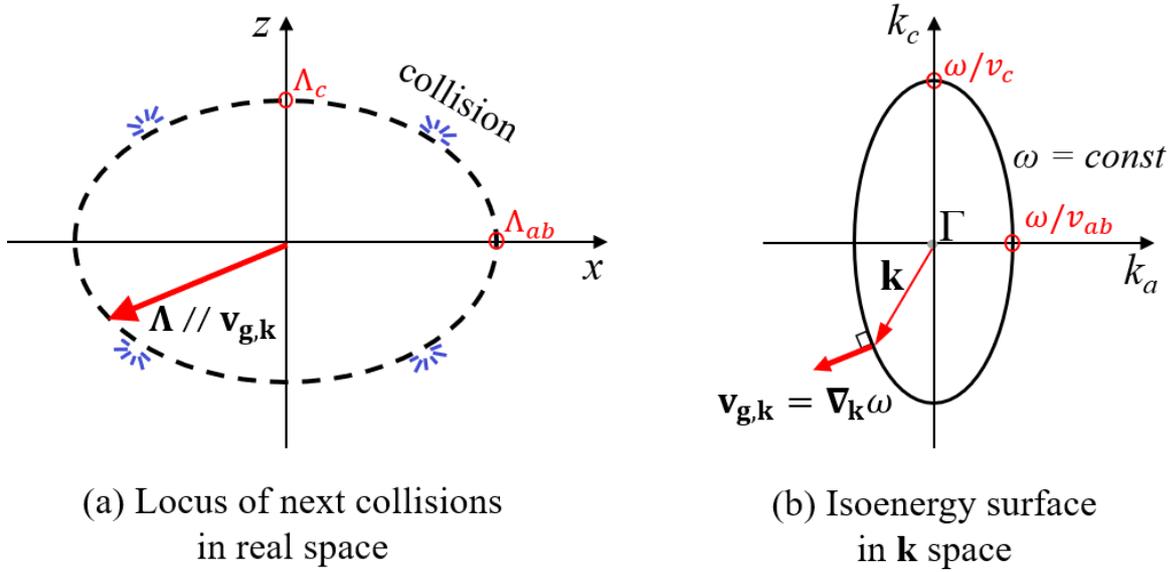

(a) Locus of next collisions in real space

(b) Isoenergy surface in **k** space

Fig. 1. (a) Key assumption: generalized imaginary locus of next possible collisions of a phonon with a specific frequency, which is initialized at the origin. The ellipsoidal surface, here projected to *x-z* plane, has an equatorial radius $\Lambda_{ab}$ and polar radius $\Lambda_c$. Here we assume $\Lambda_{ab} > \Lambda_c$; the opposite case is straightforward. (b) Isoenergy surface, here projected to $k_a - k_c$ plane, from the anisotropic Debye assumption.[10] For a specific wavevector, **k**, its group velocity is obtained by definition, $\mathbf{v_{g,k}} = \nabla_\mathbf{k}\omega$. Along this direction, the corresponding MFP is obtained using the collision surface in (a). Note here everything in this figure assumes a constant polarization.

We are now defining the model's essential phenomenological assumption which determines the MFP corresponding to a specific wavevector **k**. As concluded in the context of Eq. 1, $\tau_\mathbf{k}$ is a scalar function of **k**, and thus the direction of MFP aligns with that of the phonon group velocity $\mathbf{v_{g,k}}$, which is defined as $\mathbf{v_{g,k}} = \nabla_\mathbf{k}\omega$. For a specific **k**, along the direction of $\mathbf{v_{g,k}}$, which is perpendicular to the isoenergy surface (Fig. 1b), the MFP is obtained by drawing a line started from the origin and intercepted by the imaginary scattering surface in real space (Fig. 1a).

To proceed, here we employ the anisotropic Debye dispersion,[10,22]

$$\omega^2 = v_{ab}^2 k_{ab}^2 + v_c^2 k_c^2, \qquad (3)$$

where $v_{ab}$ and $v_c$ are the magnitudes of the sound velocity, and $k_{ab}$ and $k_c$ are the wavevectors along *ab*-plane and *c*-axis, respectively. This dispersion has ellipsoidal isoenergy surfaces (Fig. 1b).

Following the recipe above, we obtain

$$|\Lambda_{\mathbf{k}}| = \sqrt{\frac{k_{ab}^2 v_{ab}^4}{\omega^2} + \frac{k_c^2 v_c^4}{\omega^2}} \Big/ \sqrt{\frac{k_{ab}^2 v_{ab}^4}{\Lambda_{ab}^2 \omega^2} + \frac{k_c^2 v_c^4}{\Lambda_c^2 \omega^2}}. \tag{4}$$

with the corresponding relaxation time,

$$\tau_{\mathbf{k}} = |\Lambda_{\mathbf{k}}|/|\mathbf{v}_{g,\mathbf{k}}| = 1 \Big/ \sqrt{\frac{k_{ab}^2 v_{ab}^4}{\Lambda_{ab}^2 \omega^2} + \frac{k_c^2 v_c^4}{\Lambda_c^2 \omega^2}}. \tag{5}$$

Here the anisotropy of $\tau$ is apparent in its explicit $\mathbf{k}$-dependence. As discussed in detail in the supplementary material,[19] Eq. 5 results in ellipsoidal iso-scattering-rate ($\tau_{\mathbf{k}}^{-1}$) surfaces in $\mathbf{k}$ space.

As a concrete example, we extend the Klemens model[23] that describes the phonon-phonon umklapp scattering from isotropic to anisotropic. The isotropic Klemens model is traced back to Peierls[24], developed by Leibfried and Schlomann,[25] Julian,[26] and others, and finalized and popularized by Klemens.[23,27] Here we restrict the analysis to high temperatures ($T \geq 0.1 \times \max(\theta_{D,ab}, \theta_{D,c})$, where $\theta_{D,ab}$ and $\theta_{D,c}$ are the two Debye temperatures),[4,28] and thus neglect the original exponential factor that was proposed to freeze out the Umklapp process at low temperatures. Therefore, we assume that the MFP along the principal direction is

$$\Lambda_{umkl.,i}^{-1} = B_i \omega^2 T v_i^{-1}, \tag{6}$$

where $i$ denotes *ab* or *c*, and $B_i$ is a phenomenological fitting parameter that indicates the directional scattering strength. Note that Eqs. 2 and 6 define the MFP ellipsoids in Fig. 1a, and specify $\tau_{\mathbf{k}}$ (Eq. 5) that enters the $\mathbf{k}$ space integration of Eq. 1. As discussed in detail in the supplementary material,[19] this is fundamentally different from the $\nabla T$ dependent scattering model,[14,17,18] which neglects the angular dependence of $\tau_{\mathbf{k}}$, and thus causes violation of the Onsager reciprocity relations.

Analogous to the isotropic scenario,[28] one may further expect

$$B_i \propto \gamma_i^2/(v_i^2 \theta_{D,i}), \tag{7}$$

where $\gamma_i$ is the Grüneisen parameter indicating the anharmonicity of the interatomic force constant along the principal direction.[29] Therefore, while the directional sound velocity, $v_i$, is directly linked to the harmonic term of the interatomic potential, the directional scattering strength, $B_i$, is related to both the harmonic and anharmonic terms.

Substituting Eqs. 5 and 6 into Eq. 1, and incorporating the anisotropic Brillouin zone assumption,[10] $k_{ab}^2/k_{ab,m}^2 + k_c^2/k_{c,m}^2 = 1$, where $k_{ab,m}$ and $k_{ab,m}$ are the wavevector cutoffs, we obtain the thermal conductivity along the *ab*-plane and *c*-axis, respectively,

$$\kappa_{ab,umkl.} = \frac{k_{c,m}}{4\pi^2} \frac{k_B}{B_{ab}T} \sum_{pol} \int_0^1 \frac{1}{\sqrt{1-p\mu^2}} \frac{1-\mu^2}{\left(\sqrt{1-q\mu^2}\right)^3} d\mu, \tag{8a}$$

$$\kappa_{c,umkl.} = \frac{k_{ab,m}^2}{2\pi^2 k_{c,m}} \frac{k_B}{B_c T} \sum_{pol} \int_0^1 \frac{1}{\sqrt{1+\frac{p}{1-p}\mu^2}} \frac{\mu\sqrt{1-\mu^2}}{\left(\sqrt{1+\frac{q}{1-q}\mu^2}\right)^3} d\mu, \tag{8b}$$

where $p = 1 - 1/(r \cdot s)^2$ and $q = 1 - 1/r^2$. Following our previous definition,[10] $r = \theta_{D,ab}/\theta_{D,c}$ characterizes the anisotropy of the dispersion: $r > 1$ and $r < 1$ correspond to layered and chain-like materials, respectively. Here $\omega_{D,ab} = v_{ab} k_{ab,m}$ and $\omega_{D,c} = v_c k_{c,m}$ are the Debye frequencies in the *ab*-plane and along the *c*-axis, respectively. Here we further define the anisotropy of the relaxation time: $s = B_{ab}/B_c$. $s > 1$ and $s < 1$ correspond to stronger and weaker phonon-phonon scattering along the *ab*-plane than the *c*-axis, respectively. In this context, $p$ and $q$ represent the degree to which the thermal conductivity deviates from the isotropic scenario. In the isotropic case, i.e. $r = s = 1$ and thus $p = q = 0$, both Eq. 8a and 8b reduce to the well-known expression,[6,28] $\kappa_{iso.,umkl.} = \sum_{pol} \frac{1}{6\pi^2} \frac{k_D k_B}{B_{iso.}T}$, where $k_D$ is the Debye cutoff wavevector and $B_{iso.}$ is the isotropic scattering constant.

Equation 8b confirms the negative correlation between $v_{ab}$ and $\kappa_c$ in a model layered material demonstrated numerically using molecular dynamics and lattice dynamics simulations.[30] It is not so difficult to see from Eq. 8b that the increase of $v_{ab}$ (with $v_c$ fixed) leads to the decrease of both $p$ and $q$, and correspondingly the integrand and thus $\kappa_c$ decrease. Likewise, Eq. 8a indicates a similar negative correlation between $v_c$ and $\kappa_{ab}$. These negative correlations can be well understood by the phonon focusing effect.[10]

Similarly, we find from Eq. 8b that $\kappa_c$ is negatively correlated to $B_{ab}$, and thus positively correlated to $\Lambda_{ab}$; and likewise $\kappa_{ab}$ is positively correlated to $\Lambda_c$. These positive correlations between the MFP along a principal direction and the thermal conductivity along its perpendicular direction can be understood from Fig. 1a and Eq. 1. For example, increasing $\Lambda_{ab}$ elongates the imaginary collision surface along the *ab*-plane, thus increasing the MFP along all directions except directly along the *c*-axis itself, which is still pinned to $\Lambda_c$. This correspondingly increases $\kappa_c$.

These correlations between MFPs and thermal conductivities in the orthogonal directions are consistent with the observations in Ref. 30. For example, the negative correlation (Fig. 3c of Ref. 30) between the *c*-axis phonon irradiation, $H_c$, and the *ab*-plane interatomic force constant, $\chi_{ab}$, is weaker than that between $\kappa_c$ and $\chi_{ab}$ (Fig. 1c Ref. 30). This is because the former is a result of pure phonon focusing effect, while the latter also involves the negative correlation between $B_{ab}$ and $\kappa_c$. Note here $B_{ab}$ can be loosely regarded as the anharmonic part of $\chi_{ab}$.

The temperature-dependent thermal conductivities of bulk graphite along principal axes have been well documented experimentally,[31] making it a useful check of the accuracy of the anisotropic Klemens model. Our previous work shows that, without any fitting parameter, the prediction of the anisotropic Debye dispersion (Eq. 3) agrees with the experimental specific heat and the lattice dynamics calculation of phonon irradiance to within 10% throughout the temperature range 50 - 2,000 K and 200 - 10,000 K, respectively.[10]

Building upon these developments, we justify the anisotropic Klemens approximation (Eqs. 5 and 6) by comparing the thermal conductivity model (Eqs. 8a and 8b) to measurements[31] at high temperatures ($T \geq 0.1 \times \max(\theta_{D,ab}, \theta_{D,c}) \approx 200$ K, up to 2,000 K). The model accounts for contributions from the three acoustic branches, as well as the lower optical branches which have relatively high group velocities and can be viewed as continuous extension of the acoustic branches at the FBZ boundary along the *c*-axis direction[32] (e.g. TA→TO' and LA→LO' in Fig. 2 of Ref. 33). We use the same wavevector cutoffs and sound velocities as in our previous work (Table III of Ref. 10; repeated in Table S1 of the supplementary material[19] for convenience), and treat $B_{ab}$ and $B_c$ as two adjustable parameters and use a nonlinear least-squares algorithm to minimize the root mean square (rms) error of

$|\log(\kappa_{expt.}/\kappa_{model})|$,[34] where $\kappa_{expt.}$ includes both $\kappa_{ab}$ and $\kappa_c$. Note here although it is easy to calculate $\kappa_{ab}$ and $\kappa_c$ by summing over each branch, for simplicity and physical insight, we lump the three acoustic branches into one triply degenerate branch using the two constraints from our previous work (Eqs. 21 and 22a of Ref. 10: $3v_{ab,eff.}^{-2} v_{c,eff.}^{-1} = \sum_{pol} v_{ab}^{-2} v_c^{-1}$ and $3v_{ab,eff.}^{-2} = \sum_{pol} v_{ab}^{-2}$).

In Fig. 2, the experimental $\kappa_{ab}$ (circles) and $\kappa_c$ (squares)[31] in the temperature range of 200-2,000 K both show $\kappa \propto T^{-1}$ behaviors, manifesting the fact that the Umklapp phonon-phonon scattering is dominant. The best-fit model (solid red lines) corresponds to two characteristic scattering parameters $B_{ab} = 1.70 \times 10^{-20} \text{s} \cdot \text{K}^{-1}$ and $B_c = 42.5 \times 10^{-20} \text{s} \cdot \text{K}^{-1}$ (1st row of Table I), indicating much stronger phonon-phonon scattering in the $c$-axis direction than along the $ab$-plane. This is reasonable because for graphite we have $v_c < v_{ab}$, $\theta_{D,c} < \theta_{D,ab}$, and $|\gamma_c| > |\gamma_{ab}|$,[35] which supports the fitting result qualitatively according to Eq. 7.

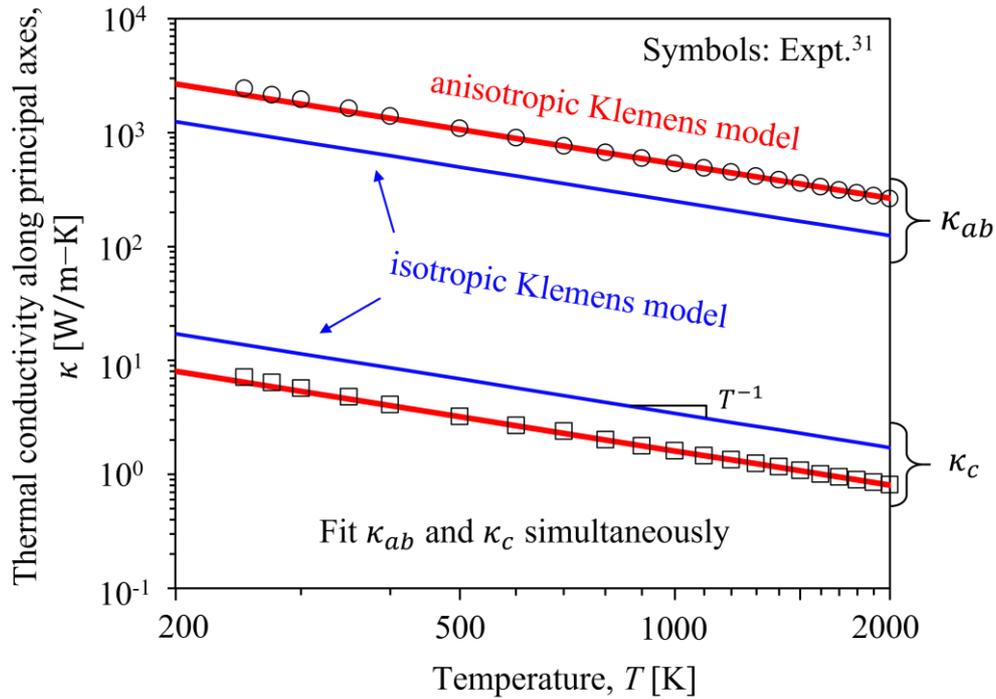

Fig. 2. Comparison with experimental data[31] for $\kappa_{ab}$ (open circles) and $\kappa_c$ (open squares) of graphite in the temperature range of 200-2,000K. Both the isotropic (solid blue line) and the anisotropic (solid red line) Klemens models capture the characteristic $T^{-1}$ power law, confirming that the phonon-phonon scattering is dominated by the Umklapp process. However, only the anisotropic model successfully reproduces $\kappa_{ab}$ and $\kappa_c$ in magnitude. Note here the best-fit results to the experimental data of both $\kappa_{ab}$ and $\kappa_c$ are obtained either by adjusting the scattering parameter $B_{iso.}$ of the isotropic model or by adjusting $B_{ab}$ and $B_c$ of the anisotropic model (Eqs. 8a and 8b). The same optimal $B_{ab}$ and $B_c$ obtained here will be used in Fig. 3 below.

For comparison, Fig. 2 also includes the best fit using the traditional isotropic Klemens model (solid blue line) with one adjustable parameter $B_{iso.} = 11.0 \times 10^{-20} \text{s} \cdot \text{K}^{-1}$ (2nd row in

Table I). The rms residual of the anisotropic scattering model (3%) is far smaller than that of the isotropic model (356%).

To reduce the rms residual of the isotropic model, one strategy is to fit $\kappa_{ab}$ with an adjustable parameter $B_{ab}$, and $\kappa_c$ with a different parameter $B_c$ (3rd row in Table 1).[14,17,18] Although this approach agrees with the experiments as good as the anisotropic Klemens model, it violates the Onsager reciprocity relations.[19] Another strategy is to consider the branch-wise scattering parameters:[16,36] instead of one single lumped parameter $B_{iso.}$ for all three acoustic branches, one may assign three adjustable parameters, $B_{TA}$, $B_{TL1}$, and $B_{TL2}$. Note here the subscripts denotes the pure transverse acoustic branch (TA) and two recomposed branches (TL1 and TL2) from decomposing the quasi-longitudinal acoustic (quasi-LA) and quasi-transverse acoustic (quasi-TA) branches.[10] This decomposition-recomposition ensures that all three branches of graphite have dispersions in the form of Eq. 3, at least in the long wavelength limit. The best fit parameters, $B_{TA} = 2.63 \times 10^{-20}$s·K$^{-1}$, $B_{TL1} = 13.6 \times 10^{-20}$s·K$^{-1}$, and $B_{TL2} = 587 \times 10^{-20}$s·K$^{-1}$ (4th row in Table I), lead to a result that is nearly indistinguishable with the anisotropic Klemens model (solid red line), and thus only the latter is shown in Fig. 2.

Table I. Best fit parameters of four models for the temperature-dependent $\kappa_{ab}$ and $\kappa_c$ of graphite (Fig. 2). All three models employ the anisotropic Debye dispersion, with the same wavevector cutoffs and sound velocities as in Table III of Ref. 10 (repeated in Table S1 of the supplementary material for convenience). For simplicity and physical insight, we lump the three acoustic branches into one triply degenerate branch for the first three models (see text for details).

| Model | Color coding in Figs. 2&3 and comments | Expression of Umklapp scattering | Best fit parameters [$10^{-20}$s·K$^{-1}$] |
|---|---|---|---|
| Anisotropic Klemens model (one lumped branch) | Red | Eqs. 5 and 6 | $B_{ab} = 1.70$ $B_c = 42.5$ |
| Isotropic Klemens model (one lumped branch) | Blue; not shown in Fig. 3 (already failed in Fig. 2) | $\tau_{iso.}^{-1} = B_{iso.}\omega^2 T$ | $B_{iso.} = 11.0$ |
| $\nabla T$ dependent Klemens model | Not shown in Figs. 2 and 3 (indistinguishable with red in Fig. 2) Violate Onsager reciprocity relations | $\tau_{ab}^{-1} = B_{ab}\omega^2 T$ $\tau_c^{-1} = B_c\omega^2 T$ | $B_{ab} = 5.14$ $B_c = 23.5$ |
| Isotropic Klemens model (branch-wise) | Green; not shown in Fig. 2 (indistinguishable with red in Fig. 2) | $\tau_{pol.}^{-1} = B_{pol.}\omega^2 T$ (*pol.* = *TA*, *TL1*, or *TL2*) | $B_{TA} = 2.63$ $B_{TL1} = 13.6$ $B_{TL2} = 587$ |

To further distinguish the two models, the lumped two-parameter anisotropic Klemens model in this work (1st row in Table I) and the branch-wise three-parameter isotropic Klemens model from literature (4rd row),[16,36] we consider the size effect on $\kappa_c$ of graphite. We assume perfectly thermalizing (black) boundary conditions: $I_\mathbf{k}^+(z=0) = \frac{1}{8\pi^3}\hbar\omega|\mathbf{v}_{\mathbf{g},\mathbf{k}}|f_{BE}(T_1)$ and $I_\mathbf{k}^-(z=L) = \frac{1}{8\pi^3}\hbar\omega|\mathbf{v}_{\mathbf{g},\mathbf{k}}|f_{BE}(T_2)$, where $f_{BE}$ is the Bose-Einstein distribution, and $T_1$ and $T_2$ are the emissive temperature of the two boundaries.

We start from the net heat flux along the *c*-axis direction,[10]

$$q_{net,c} = \sum_{pol} \iiint I_\mathbf{k} \hat{\mathbf{s}} \cdot \hat{\mathbf{c}} d^3\mathbf{k}, \qquad (9)$$

where $\hat{\mathbf{c}}$ is the unit vector along the $c$-axis, $\hat{\mathbf{s}}$ is a unit vector parallel to $\mathbf{v}_{g,\mathbf{k}}$, and $I_{\mathbf{k}} = \frac{1}{8\pi^3}\hbar\omega|\mathbf{v}_{g,\mathbf{k}}|f$ is the phonon intensity,[10] in which $f$ is the distribution function. Note here the integration runs over the $4\pi$ solid angle. We evaluate $I_{\mathbf{k}}$ and $q_{net,c}$ by solving the steady state equation of phonon radiative transfer under the relaxation time approximation,

$$\cos\theta \frac{\partial I_{\mathbf{k}}}{\partial z} = \frac{I_{0,\mathbf{k}} - I_{\mathbf{k}}}{|\Lambda_{\mathbf{k}}|}, \qquad (10)$$

where $z$-axis and $c$-axis are aligned, $\theta$ is the angle measured from $\hat{\mathbf{c}}$ to $\hat{\mathbf{s}}$, and the equivalent equilibrium phonon intensity, $I_{0,\mathbf{k}}$, can be obtained by averaging the intensity of all phonons around a local point.[21,37] Following Ref. 20 we numerically solve Eq. 10 using the two-flux model,[20]

$$I_{\mathbf{k}}(z,\theta) = \begin{cases} I_{\mathbf{k}}^+(z,\theta), & 0 < \theta < \pi/2 \\ I_{\mathbf{k}}^-(z,\theta), & \pi/2 < \theta < \pi \end{cases}. \qquad (11)$$

Here the information of phonon dispersion and phonon-phonon scattering enters the calculation through $|\Lambda_{\mathbf{k}}|$ in Eq. 10. Finally, the thermal conductivity along $c$-axis is obtained from $\kappa_c = q_{net,c}L/(T_1 - T_2)$.

Figure 3 compares recent measurements (circles) by Fu et al.[38] for thickness dependent $\kappa_c$ of graphite at 300 K to the the three-parameter branch-wise isotropic Klemens model[16,36] (solid green line) and our lumped two-parameter anisotropic Klemens model (solid red line), all normalized to the corresponding bulk values.[39] Note that the rms residuals of the two models fitting to the experimental results of $\kappa_{ab}$ and $\kappa_c$ of the bulk graphite sample in Fig. 2 are both 3%. With no further adjusting these best fit parameters, however, our lumped two-parameter anisotropic Klemens model (solid red line) agrees much better with the measurements of the thickness dependent $\kappa_c$ (circles). This is because the branch-wise three-parameter isotropic Klemens model (solid green line) is equivalent to averaging out the longer $\Lambda_{ab}$ and the shorter $\Lambda_c$ to obtain an $\Lambda_{iso.}$ that is longer than the actual $\Lambda_c$. This overestimate of the $c$-axis MFP leads to stronger size effect, and thus smaller $\kappa_c$ of graphite films. This interpretation is further confirmed in the supplementary material by the thermal conductivity accumulation function calculation.[19]

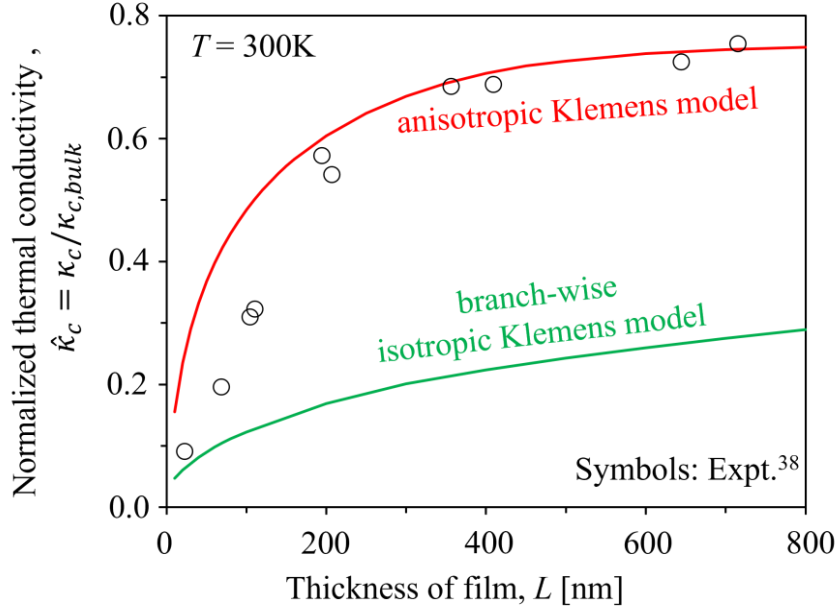

Fig. 3. Comparison with experimental data[38] (points) for normalized $\kappa_c$ of graphite thin films at 300 K. Although rms residues to the temperature dependent $\kappa_{ab}$ and $\kappa_c$ (Fig. 2) of the two models are both 3%, here the anisotropic Klemens model (solid red line) with two adjustable parameters, $B_{ab}$ and $B_c$, of this work agrees with the experiment much better than the branch-wise isotropic Klemens model[16,36] (solid green line) with three adjustable parameters, $B_{TA}$, $B_{TL1}$, and $B_{TL2}$.

In summary, we have proposed an anisotropic Klemens model for phonon-phonon Umklapp scattering. Combined with the anisotropic Debye dispersion, this anisotropic scattering model underlines the constraints on the thermal conductivity tensor imposed by the Onsager reciprocity relations, and offers analytical expressions for both $\kappa_{ab}$ and $\kappa_c$, suitable for both layered and chainlike materials with any anisotropy ratio of the dispersion and the scattering. When compared to experimental data of graphite, this framework not only fits the temperature dependent $\kappa_{ab}$ and $\kappa_c$ between 200-2,000K, but also is consistent with the thickness dependent $\kappa_c$ at 300 K. In contrast, the isotropic scattering model with either one lumped branch or three branches fails in interpreting at least one set of experiment above.

This work was supported in part by the National Natural Science Foundation of China (51776038). We thank Hengrui Chen and Zhiyong Wei for helpful discussions, and Dongyan Xu for sharing their original experimental data in Ref. 38. In particular, we thank Chris Dames for critically reading this manuscript with many constructive comments and suggestions.

Supplementary Material

**Table of Contents:**
A. Constraint on the relaxation time imposed by Onsager reciprocity relations
B. The arbitrary anisotropic thermal conductivity tensor
C. Expressions of the nine entries of the 3 by 3 matrix, $\bar{\bar{\kappa}}^{xyz}$, of Eq. S4
D. Iso-relaxation-time surfaces
E. input parameters for graphite
F. Thermal conductivity accumulation function along the $c$-axis

**A. Constraint on the relaxation time imposed by Onsager reciprocity relations**

A straightforward and seemingly quite reasonable, phenomenological approach to model the direction-dependent thermal conductivities is to assign temperature-gradient-direction dependent ($\boldsymbol{\nabla} T$ dependent) relaxation times [S1-S3], which modifies Eq. 1 of the main text to

$$\kappa_{\alpha\beta} = \sum_{pol}\left[\sum_{\mathbf{k}} C_{\mathbf{k}} \cdot (\mathbf{v}_{g,\mathbf{k}} \cdot \hat{\alpha}) \cdot (\mathbf{v}_{g,\mathbf{k}} \cdot \hat{\beta}) \cdot \tau_{\beta,|\mathbf{k}|}\right], \quad (S1)$$

with $\tau_{\mathbf{k}}$ replaced by $\tau_{\beta,|\mathbf{k}|}$. Recall here $\alpha$ and $\beta$ denote the direction of the heat flux and the temperature gradient, respectively.

Although seemingly similar to Eq. 1 of the main text, Eq. S1 here can violate the well-established symmetry of the thermal conductivity tensor: $\kappa_{\alpha\beta} = \kappa_{\beta\alpha}$ (known as Onsager reciprocity relations [S4]). However, here if $\alpha$ and $\beta$ are not principal directions and if $\alpha \neq \beta$, Eq. S1 results in $\kappa_{\beta\alpha} \neq \kappa_{\alpha\beta}$, since $\tau_{\alpha,\mathbf{k}}$ does not necessarily equal $\tau_{\beta,\mathbf{k}}$. Therefore, we conclude that the Onsager reciprocal relations impose a fundamental constraint on $\tau_{\mathbf{k}}$, and forbids its $\boldsymbol{\nabla} T$ dependence.

Note here we restrict the analysis to the framework of the Boltzmann Transport Equation (BTE) under the Relaxation Time Approximation (RTA), and thus this $\boldsymbol{\nabla} T$ dependent scattering time (Eq. S1) is fundamentally different from the effective scattering time defined from the iterative solution of the inelastic BTE in Ref. S5.

Figure S1 further distinguishes the prior $\boldsymbol{\nabla} T$ dependent scattering model [S1-S3] from the anisotropic scattering model (Eqs. 5 and 6 of the main text) proposed in this work. With the anisotropic Debye dispersion [S6], here we use the dots on the isoenergy surfaces to represent phonon modes with wavevector $\mathbf{k}$, and let the size of the dots represent the magnitude of the relaxation time of that microstate. Note here, to be consistent with the description in the main text, we drop the polarization dependence of the phonon mode. In the $\boldsymbol{\nabla} T$ dependent scattering model (Fig. S1a), when $\boldsymbol{\nabla} T$ is in the $ab$-plane, the relaxation time of the microstates on a specific isoenergy surface is all the same, which is written as $\tau_{ab,|\mathbf{k}|}$; likewise, when $\boldsymbol{\nabla} T$ is along the $c$-axis, the relaxation time of the microstates on the same isoenergy surface is also the same, which is written as $\tau_{c,|\mathbf{k}|}$. But here $\tau_{ab,|\mathbf{k}|} \neq \tau_{c,|\mathbf{k}|}$. In contrast, the anisotropic scattering model leads to different relaxation times even for the microstates on the same isoenergy surface. These relaxation times will not change according to the direction of $\boldsymbol{\nabla} T$.

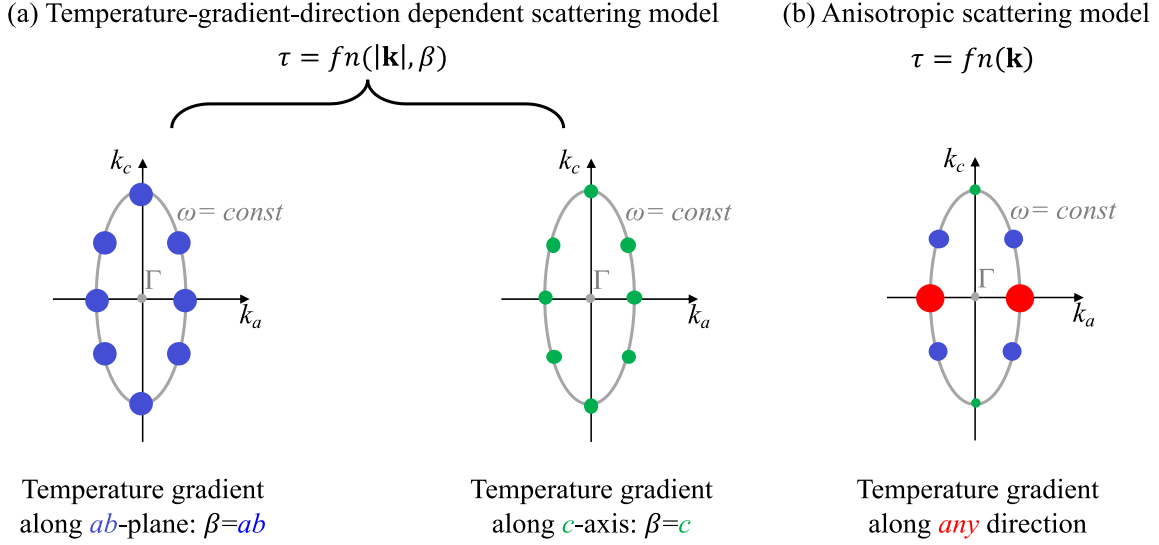

Fig. S1. Visualization of the distinction between (a) the $\nabla T$ dependent scattering model in Refs. S1-S3 and (b) the anisotropic scattering model of this work. Here we assume anisotropic Debye dispersion with ellipsoidal isoenergy surfaces (see Fig. 1b of the main text [S6]). The dots represent phonon modes with wavevector, **k**, and the size of the dots represents the magnitude of the relaxation time.

## B. The arbitrary anisotropic thermal conductivity tensor (of materials with hexagonal symmetry)

We now show $\kappa_{\alpha\beta}$ in Eq. 1 of the main text, derived from BTE under RTA, can be expressed as a linear combination of the principal components of the conductivity tensor: $\kappa_{ab}$ and $\kappa_c$, which are obtained from Eqs. 8a and 8b of the main text. Substituting $\mathbf{v_{g,k}} = (v_{g,a}, v_{g,b}, v_{g,c})$, and a pair of orthogonal unit vectors, $\hat{\alpha} = (\alpha_1, \alpha_2, \alpha_3)$ and $\hat{\beta} = (\beta_1, \beta_2, \beta_3)$, into Eq. 1, one obtains

$$\kappa_{\alpha\beta} = \alpha_1\beta_1\kappa_{ab} + \alpha_2\beta_2\kappa_{ab} + \alpha_3\beta_3\kappa_c, \tag{S2}$$

where $\kappa_{ab} = \sum_{pol}\left[\sum_{\mathbf{k}} C_{\mathbf{k}} \cdot v_{g,a}^2 \cdot \tau_{\mathbf{k}}\right] = \sum_{pol}\left[\sum_{\mathbf{k}} C_{\mathbf{k}} \cdot v_{g,b}^2 \cdot \tau_{\mathbf{k}}\right]$, and $\kappa_c = \sum_{pol}\left[\sum_{\mathbf{k}} C_{\mathbf{k}} \cdot v_{g,c}^2 \cdot \tau_{\mathbf{k}}\right]$. Here the $\theta$'s are defined in Fig. S2. Note here we utilized the symmetry of this *a-b-c* coordinate with respect to the FBZ to obtain $\sum_{\mathbf{k}} C_{\mathbf{k}} \cdot v_{g,i} \cdot v_{g,j} \cdot \tau_{\mathbf{k}} = 0$, where $i, j = a, b,$ or $c$, and $i \neq j$.

Equation S2 can be confirmed using the classic tensor transformation [S7] in the framework of Fourier's law. In the principal *a-b-c* coordinate (Fig. S2), the thermal conductivity tensor reads

$$\bar{\bar{\kappa}}^{abc} = \begin{bmatrix} \kappa_{ab} & 0 & 0 \\ 0 & \kappa_{ab} & 0 \\ 0 & 0 & \kappa_c \end{bmatrix}. \tag{S3}$$

Using the classic tensor rotation rule [pg. 158 of Ref. S7], one obtains the conductivity tensor in an arbitrary *x-y-z* coordinate (Fig. S2),

$$\bar{\bar{\kappa}}^{xyz} = \bar{\bar{R}}\bar{\bar{\kappa}}^{abc}\bar{\bar{R}}^T, \tag{S4}$$

where the rotation tensor,

$$\bar{\bar{R}} = \begin{bmatrix} \cos\theta_{xa} & \cos\theta_{xb} & \cos\theta_{xc} \\ \cos\theta_{ya} & \cos\theta_{yb} & \cos\theta_{yc} \\ \cos\theta_{za} & \cos\theta_{zb} & \cos\theta_{zc} \end{bmatrix} \tag{S5}$$

bridges the two sets of coordinates: *a-b-c* and *x-y-z*. We outline the detailed expression of nine entries of the 3 by 3 matrix $\bar{\bar{\kappa}}^{xyz}$ (Eq. S4) in Appendix D.

The BTE result (Eq. S2) recovers the classic Fourier's law result (Eq. S4). For example, if $\hat{\alpha}$ is aligned with *x*-axis are $\hat{\beta}$ is aligned with *z*-axis, i.e. $\hat{\alpha} = (\alpha_1, \alpha_2, \alpha_3) = (\cos\theta_{xa}, \cos\theta_{xb}, \cos\theta_{xc})$ and $\hat{\beta} = (\beta_1, \beta_2, \beta_3) = (\cos\theta_{za}, \cos\theta_{zb}, \cos\theta_{zc})$, one immediately obtains $\kappa_{\alpha\beta} = \bar{\bar{\kappa}}^{xyz}_{xz} = \cos\theta_{xa}\cos\theta_{za}\kappa_{ab} + \cos\theta_{xb}\cos\theta_{zb}\kappa_{ab} + \cos\theta_{xc}\cos\theta_{zc}\kappa_c$.

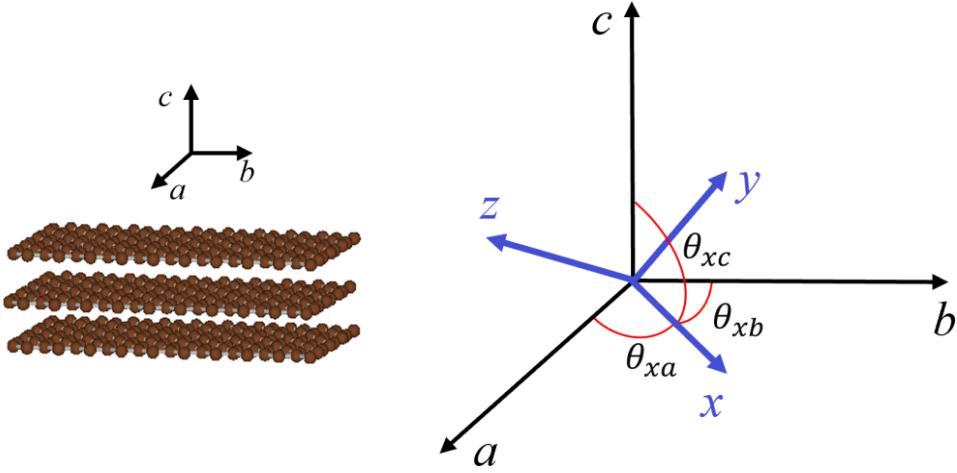

Figure S2. An arbitrary *x-y-z* coordinate rotated from the principal *a-b-c* coordinate. For clarity, only the angles between x-axis and the *a-b-c* axes are labeled.

The analyses above emphasize that once any two entries of the $\bar{\bar{\kappa}}^{xyz}$ matrix are compared to the corresponding experimental data to obtain the two adjustable parameters, $B_{ab}$ and $B_c$, all the other entries of the $\bar{\bar{\kappa}}^{xyz}$ matrix under any arbitrary orientation can be obtained.

For the case of the $\nabla T$ dependent scattering time model [S1-S3] (Eq. S1), however, the tensor rotation (Eq. S4) cannot be applied to obtain the arbitrary anisotropic thermal conductivity tensor, since an unsymmetrical tensor (Eq. S1) *cannot* be diagonalized to the form of Eq. S3 by any series of rotating transformations [S8]. Instead, fits of *B*'s have to be re-performed for each orientation. Figure S3a shows a specific example of such fitting, in which the *x*-axis is aligned with the principal *a*-axis while *y*- (and *z*-) axis rotates an angle $\theta = 30°$ in a counterclockwise direction (inset of Fig. S3b). Here we synthetize "experimental" data (points in Fig. S3a) using the anisotropic Klemens model of this work and the classic tensor rotation rule (Eq. S4), and adjust $B_{yy}(\theta = 30°)$ for the best fit of $\kappa_{yy}$, and $B_{zz}(\theta = 30°)$ for the best fit of $\kappa_{zz}$, respectively. Figure S3b summarizes the successive fitting results of $B_{yy}$ and $B_{zz}$ as a function of the rotation angle, $\theta$. Figure S3 shows the complexity to implement the $\nabla T$ dependent scattering model (Eq. S1) to arbitrary oriented conductivity tensor, even if one omits the fact that this model violates the Onsager reciprocity relations.

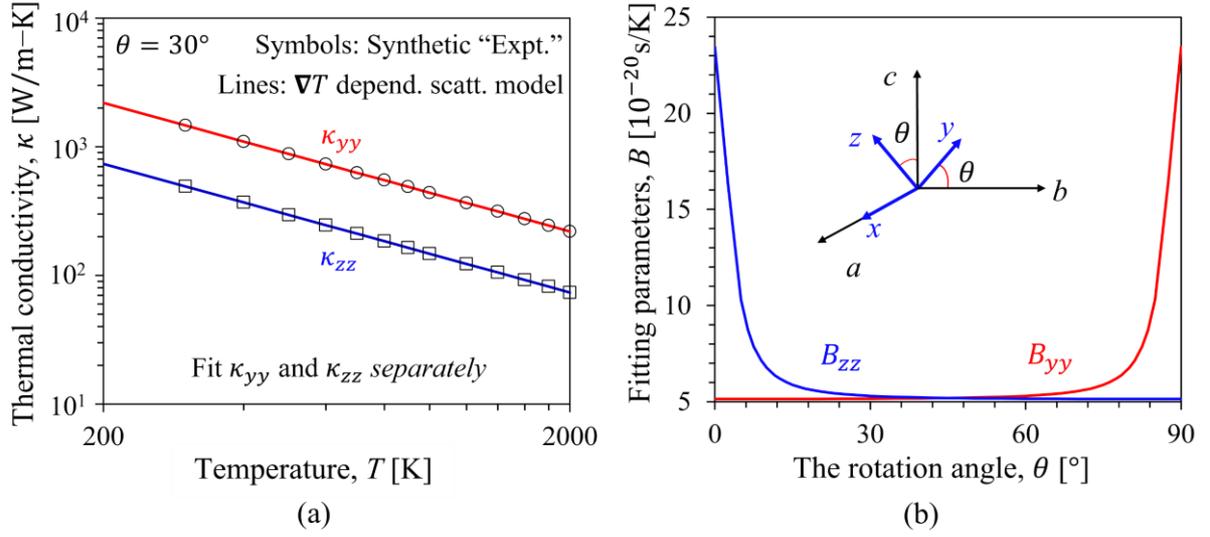

Figure S3. Illustration of the complexity of implementing the $\nabla T$ dependent scattering model (Eq. S1): fittings have to be performed for each arbitrary angle. (a) Best fit (lines) using the $\nabla T$ dependent scattering model [S1-S3] to "experimental" data of $\kappa_{yy}$ and $\kappa_{zz}$ (points) synthesized using the anisotropic Klemens model of this work and the classic tensor rotation rule for a specific angle, $\theta = 30°$. (b) Successive fitting results of $B_{yy}$ and $B_{zz}$ as a function of the rotation angle, $\theta$.

## C. Expressions of the nine entries of the 3 by 3 matrix, $\bar{\bar{\kappa}}^{xyz}$, of Eq. S4

$$\begin{aligned}
\bar{\bar{\kappa}}^{xyz}_{xx} &= (\cos^2\theta_{xa} a_{11}^2 + \cos^2\theta_{xb} a_{12}^2)\kappa_{ab} + \cos^2\theta_{xc} a_{13}^2 \kappa_c \\
\bar{\bar{\kappa}}^{xyz}_{yy} &= (\cos^2\theta_{ya} + \cos^2\theta_{yb})\kappa_{ab} + \cos^2\theta_{yc} = \kappa_c \\
\bar{\bar{\kappa}}^{xyz}_{zz} &= (\cos^2\theta_{za} + \cos^2\theta_{zb})\kappa_{ab} + \cos^2\theta_{zc}\kappa_c \\
\bar{\bar{\kappa}}^{xyz}_{xy} &= \bar{\bar{\kappa}}^{xyz}_{yx} = (\cos\theta_{xa}\cos\theta_{ya} + \cos\theta_{xb}\cos\theta_{yb})\kappa_{ab} + \cos\theta_{xc}\cos\theta_{yc}\kappa_c \\
\bar{\bar{\kappa}}^{xyz}_{xz} &= \bar{\bar{\kappa}}^{xyz}_{zx} = (\cos\theta_{xa}\cos\theta_{za} + \cos\theta_{xb}\cos\theta_{zb})\kappa_{ab} + \cos\theta_{xc}\cos\theta_{zc}\kappa_c \\
\bar{\bar{\kappa}}^{xyz}_{yz} &= \bar{\bar{\kappa}}^{xyz}_{zy} = (\cos\theta_{ya}\cos\theta_{za} + \cos\theta_{yb}\cos\theta_{zb})\kappa_{ab} + \cos\theta_{yc}\cos\theta_{zc}\kappa_c
\end{aligned} \quad (S6)$$

## D. Iso-relaxation-time surfaces

Equation 5 of the main text can be re-arranged to be

$$\left(\frac{\omega}{\tau_k/\tau_c}\right)^2 = s^2 v_{ab}^2 k_{ab}^2 + v_c^2 k_c^2, \tag{S7}$$

where $s = \frac{\tau_c}{\tau_{ab}} = \frac{\Lambda_c/v_c}{\Lambda_{ab}/v_{ab}} = \frac{B_{ab}}{B_c}$, in which the last equal sign holds for the anisotropic Klemens model (Eq. 6 of the main text). Note here the subscript, *ab* or *c*, indicates that the wavevector **k** points to $k_{ab}$ or $k_c$ axis, respectively. Apparently Eq. S7 represents ellipsoidal iso-relaxation-time surfaces in **k** space for fixed phonon frequency, $\omega$, and polarization, which, as compared to the iso-energy surfaces, is elongated (suppressed) along the $k_{ab}$ axis for $s < 1$ ($> 1$), as shown in Fig. S4.

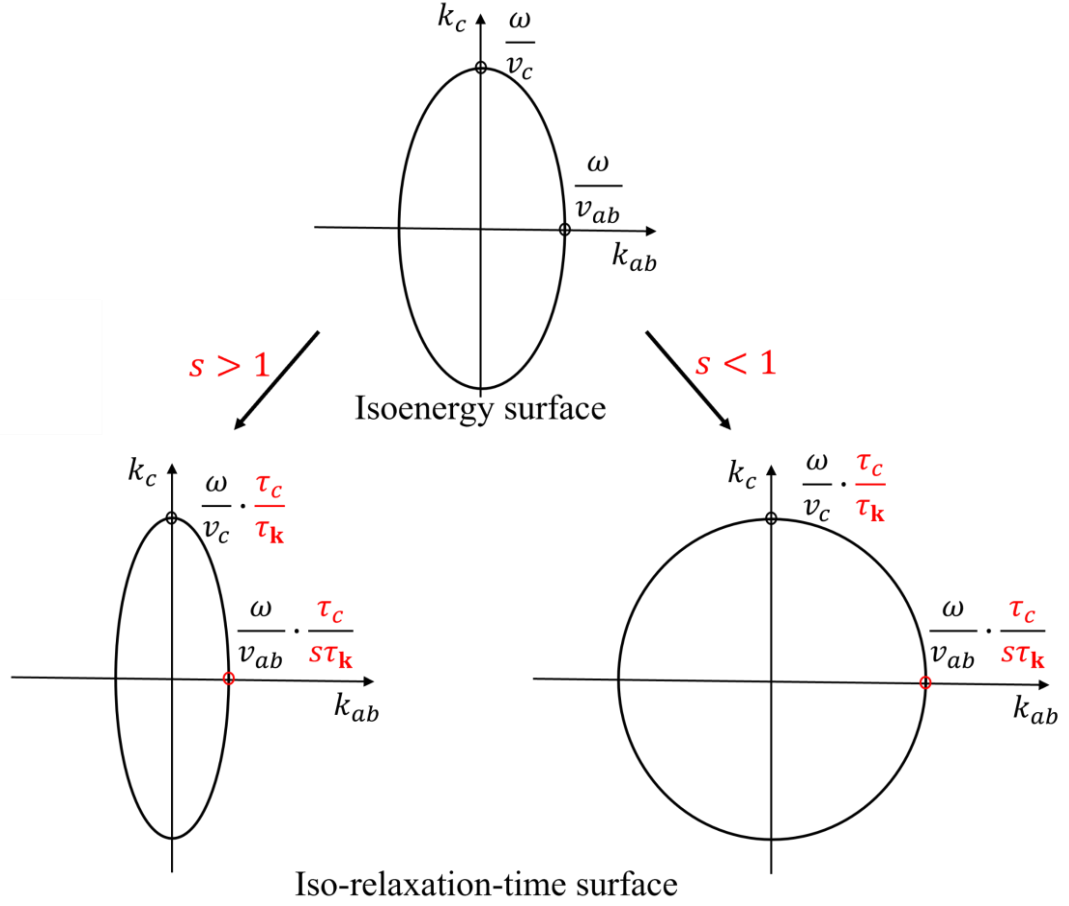

Figure S4. Iso-energy surface vs. Iso-relaxation-time surface (for fixed phonon frequency and polarization). Note here $s = \frac{\tau_c}{\tau_{ab}} = \frac{\Lambda_c/v_c}{\Lambda_{ab}/v_{ab}}$, in which the subscript, *ab* or *c*, indicates that the wavevector **k** points to $k_{ab}$ or $k_c$ axis, respectively.

### E. input parameters for graphite

Table S1. The input parameters for graphite in Ref. S6, repeated here for convenience.

| Parameter | Unit | *ab* plane | *c* axis |
|---|---|---|---|
| $v_{TA}$ | m/s | 10200 | 1000 |
| $v_{TL1}$ | m/s | 16200 | 1000 |
| $v_{TL2}$ | m/s | 6400 | 2500 |
| $k_{max}$ | $10^{10}$ m$^{-1}$ | 1.73 | 1.1 |

### F. Thermal conductivity accumulation function along the *c*-axis

To further confirm our interpretation to Fig. 3 of the main text, we compute the accumulation function [S9, S10] of $\kappa_c$. Figure S5 shows that phonons with $\Lambda_c$ shorter than 714 nm, the thickest sample of Fu's measurements [S11], contribute more than 85% to $\kappa_c$, according to our lumped two-parameter anisotropic Klemens model (solid red line); however, based on the branch-wise three-parameter isotropic Klemens model [S12, S13] (solid green line), these same phonons contribute less than 45%. As a result, for this sample, the branch-wise model predicts that those phonons with $\Lambda_c$ longer than 714 nm, which contribute more than 55% to $\kappa_c$, will be heavily truncated, leading to lower $\kappa_c$

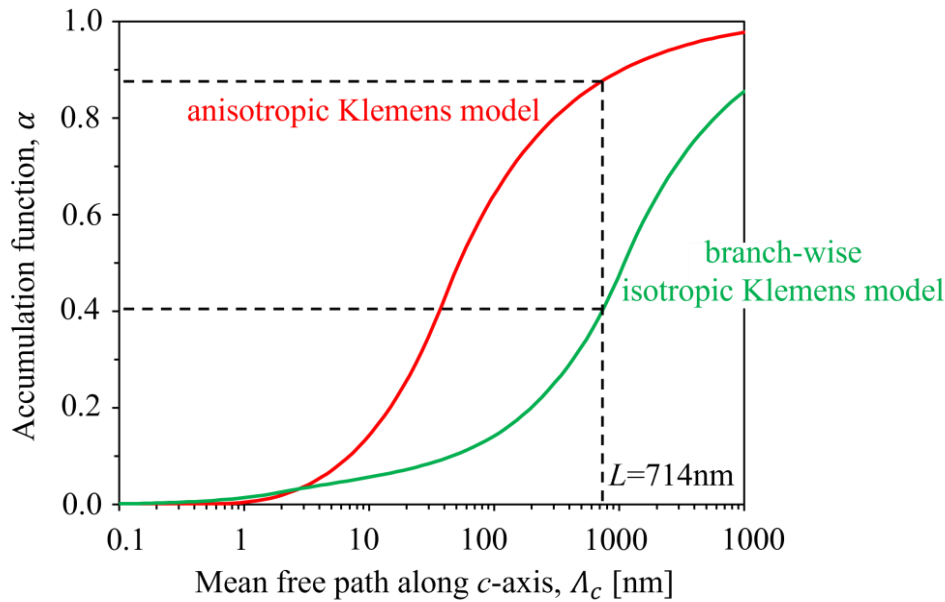

Figure S5. Accumulation functions according to the anisotropic Klemens model of this work and the branch-wise isotropic Klemens from literature [S12, S13], showing that the latter puts more weightings on the long MFP phonons and thus leads to stronger size effect on $\kappa_c$.